# STATUS REPORT OF THE INTER-LABORATORY TASK FORCE ON REMOTE OPERATION

Paul Czarapata, FNAL; Don Hartill, Cornell; Steve Myers, CERN; Stephen Peggs, BNL
Nan Phinney, SLAC; Mario Serio, INFN; Nobu Toge, KEK; Ferdinand Willeke, DESY
Chuan Zhang , IHEP Beijing

Stanford Linear Accelerator Center
Stanford University
Stanford, CA  94309

**Report of the ICFA Taskforce on Technical Aspects of a Global Accelerator Network**

# Status Report of the Inter-Laboratory Task Force on Remote Operation

**November 2001**

Paul Czarapata, FNAL; Don Hartill, Cornell; Steve Myers, CERN; Stephen Peggs, BNL; Nan Phinney, SLAC; Mario Serio, INFN; Nobu Toge, KEK; Ferdinand Willeke, DESY; Chuan Zhang , IHEP Beijing

**Executive Summary**

In February 2000, the International Committee for Future Accelerators initiated a study of a new model for international collaboration on a future large accelerator project, the Global Accelerator Network [1]. The study is based on a model of a facility, which is remote from most of the collaborating institutions. It is designed, built and operated by a collaboration of equal partner institutions distributed around the world. According to this model, the expert-staff from each laboratory remains based at their home institution but continues to participate in the operation of the machine after construction. This report summarizes the conclusions of the Task Force on Remote Operation, which investigated the general and technical implications of far-remote operations.

The task force considered the full range of activities involved in the operation of a complex accelerator including commissioning, normal operation, machine development, maintenance, troubleshooting and repair. As far as maintenance, troubleshooting and repair is concerned, the experience from existing laboratories is encouraging. It indicates that most of these activities are already performed 'remotely', or could be with properly designed equipment. The experts are required to be physically present only during initial commissioning of the hardware and for troubleshooting particularly difficult problems. Repairs require a local technically trained maintenance crew but even for the complex RF and power supply systems at HERA and at LEP/SPS, 90-95% of the interventions are made without consulting an expert and most of the rest are resolved with only a phone call. Only a few times a year is expert presence required. If one takes into account this experience for a future large accelerator facility, one may conclude that it should be possible to perform most of the maintenance, troubleshooting and repair remotely. This, however, requires comprehensive remote diagnostics, modular design of components, and a high level of standardization.

A particular question was whether an accelerator could be operated remotely, possibly even from a control room on a different continent. Most of presently operated accelerators have modern control systems which use a layered approach where high speed, high bandwidth, closed loop control is all performed locally, including time-critical machine protection functions. Console applications provide control, monitoring and diagnostics and require a slower data rate commensurate with human response times. At many sites, the consoles are 'remote' from the actual control computers and at SLAC, for example, consoles may be run from office or home. The most significant bandwidth demand in a modern control room is for real-time signals, which are used for continuous monitoring by the operations staff. In older installations, these are frequently analog signals but with advancing technology, essentially all diagnostic devices used to generate such real-time signals now have digital output and network access.

Taking this into consideration, we conclude that there appears to be no technical obstacle to far-remote control of an accelerator. Nonetheless, a dedicated high-speed network connection to the remote control room would possibly be required to supply sufficient guaranteed bandwidth for this real-time data. The rapid rate of development of communications technology should easily support the demands of accelerator operation in 5-10 years.

Staffing of the accelerator site is an important question. One has to distinguish between normal operation and shut down period. We expect that a future facility, like most accelerators, has a yearly schedule which includes something like 9 months of operation and 3 months of shutdown for maintenance and installation of upgrades.

During the period of normal operation, the local staff at the accelerator site proper is expected to be much smaller than a large laboratory typically keeps available for its facilities. A reliable number of the minimum staff necessary depends naturally very much on the details of the remote facility but experience from large machines such as HERA indicates that is could be as small as 100-200 people [2]. Staff is required for maintenance and security, for radiation safety, as well as a small non-expert intervention crew for maintenance of the accelerator hardware, exchanging components, and other minor repairs. Some of these functions could be performed by local contractors. Highly specialized tasks such as klystron replacement or vacuum intervention would require a small team of experts either on site or in close proximity. Staff would also be required for maintenance and repair of the experiments.

The extended shutdown periods are typically periods of intense activity with a large number of experts involved in installation, maintenance or upgrade activities. A remote site would have to provide sufficient infrastructure to efficiently support the expanded on-site presence.

The major challenge of such a facility lies in solving the complex management, sociological and communication problems. In designing a management and oversight structure, much can be learned from the large high-energy physics experiments and modern astronomy projects which typically involve international collaborations of distant institutions. Tight coordination is required in all phases of design, construction and installation. Operation of the accelerator is not an easy task. Energy frontier facilities are inevitably pushing the limits of accelerator technology and present unanticipated difficulties, which require intense effort from a dedicated team of experts to diagnose and solve each new problem. Past experience has shown how critical it is for these experts to have offices near each other to facilitate exchange of ideas and information. Equally important is contact between the experimenters and the accelerator physicists, and between the physicists, engineers and operations staff. To encourage an effective interchange between these disparate groups, it will be necessary to have a critical mass of experts located in at least one of the laboratories. Even then, some decreased efficiency in operation or in responding to a crisis will be inevitable because of the difficulties in communication. Many of these problems would be ameliorated if the facility were located adjacent to one of the collaborating laboratories.

# 1 Introduction

The next generation of particle accelerators will be major projects which may require a new mode of international and inter-laboratory collaboration. They are likely to be too costly to be funded by a single nation and too large to be built by a single laboratory. The tremendous technical challenge of a new facility requires a critical mass of highly qualified and experienced physicists and engineers. These experts are presently distributed among the major accelerator centers around the world and it is believed important to maintain and develop this broad base of expertise. The successful accelerator technology development of recent decades depended on extensive exchange of people with complementary technical skills. Therefore, it is desirable and probably necessary that several accelerator laboratories will participate in any future project. A consequence of a multi-laboratory project is that the accelerator will be located a considerable distance from most of the contributing institutions which design, build and operate it.

These considerations led the International Committee for Future Accelerators to initiate a study on the general and technical implications of such a collaboration. Two task forces were formed in February 2000 to conduct this study and they were asked to prepare a report on a time scale of one year. The task force on Remote Operation included members from most of the major accelerator laboratories around the world with expertise on accelerator operation, controls software, communication technologies, hardware design and maintenance. The task force members gathered information from the experts at their own institutions and from available experience in other fields, particularly astronomy.

The task force on Remote Operations began by developing a model for an international multi-laboratory collaboration to construct and operate an accelerator facility. This model is described in section 3. While it is clear that there are numerous alternative scenarios, the model was intended to provide a structure for addressing the most important technical and sociological issues. In particular, the task force attempted to answer the following questions:

- How much of the activities of commissioning, operation, machine development, maintenance, troubleshooting and repair can or should be done remotely?
- What local staff is required for operations, maintenance, or repair?
- What needs to be changed or added to the design of the hardware components to allow remote diagnosis and analysis? Are the costs of these changes significant?
- What are the requirements on the control system data transmission speed or bandwidth to support remote operation? Are presently available communication technologies a limitation requiring further R&D?
- What are the sociological problems and aspects of worldwide decentralized operation? What is required for effective communication of experience, data, parameters, ideas to allow for an adequate discussion of the problems expected during commissioning, tune-up, failure analysis, performance and reliability improvements?
- What new technical tools must be developed to ease operation of a remote facility?

This document summarizes the conclusions of the task force on Remote Operation.

## 2 Relevant Experience on Remote Operations

Existing large accelerators such as LEP and HERA, as well as smaller projects such as SLC and PEP-II, are essentially remotely operated facilities where the control system architecture supports 'far-remote' operation. Feedback loops which require fast response are implemented as locally as possible and do not require continuous intervention from the main control room by operators or console application software. Analog signals today are almost always digitized before transmission to the control room so that there is no loss of information through long cable runs. The enormous advances in computing and networking have made digital media the most convenient and inexpensive method for transmitting data, even over short distances.

The large size of present accelerators and the limited access demands that interventions be well planned and undertaken only after they have been diagnosed remotely to the largest extent possible. At some large facilities such as LEP and HERA, non-expert maintenance or repair personnel are able to handle the majority of failures and repairs. In difficult cases, they are assisted by experts via telephone or via remote computer access to the components. The unscheduled presence of experts on site is a rare exception. Detailed reports and analysis from LEP and HERA which support these conclusions are available as appendices to this document.

The commissioning, operation and optimization of the SLC is perhaps the most relevant experience for a future linear collider. However, because the SLC was an upgrade of the existing SLAC linac, many of the technical components were not modern enough to support remote maintenance and troubleshooting. A significant presence of expert staff on site was required. The control system was designed to allow consoles to be run remotely from home or office. With proper coordination, they could be run from other laboratories. However, operators in the SLC control center relied on many analog signals from older diagnostics which were not available remotely. In addition, although extensive feedback systems were developed for the SLC to stabilize the beam parameters and even optimize luminosity, some tasks still required frequent operator action with a rather fast response via the control links into the control room. This experience might seem discouraging for the feasibility of far-remote operations, but none of these technical limitations are fundamental given modern technology. The steady increase in SLC performance was often enabled by the good data logging systems and the possibility of offline analysis. Such analysis could have been performed from anywhere in the world provided the data were available to a strong, motivated external group. In fact, many aspects of the SLC experience with feedback, automated procedures and complex analysis are encouraging for far-remote operation.

Non-accelerator projects also have extensive experience with remote operation of complex technical systems with restricted accessibility. The successful operation of space experiments as well as operation of distant telescopes demonstrates that efficient remote operation is possible, practicable and routinely performed. In particular, many observatories are built in rather inhospitable locations and operated with only very little technical support on site. Troubleshooting and consultation with experts is almost exclusively performed remotely. The European space agency ESO has remotely operated telescopes in Chile from a control center in Germany for more than a decade. Their operational experience is encouraging but demonstrates that the main difficulties lie in communication and organization when handling exceptional situations and emergencies. These institutions maintain a strong presence of experts on site despite the unfavorable conditions in order to mitigate these problems. The collaborators on a remote accelerator project should carefully analyze and learn from the ESO experience.

## 3 Model of an International Multi-laboratory Accelerator Project

In order to address the issue of remote operations properly, a model of a remotely operated facility has been developed. Some of the overall management and organizational aspects relevant for remote operations are described only briefly below while further details remain to be worked out.

### 3.1 Organizational Framework for Remote Operations

The accelerator would be built and operated by a consortium of institutes, laboratories or groups of laboratories, referred to as collaborators. Each collaborator is responsible for a complete section of the machine including all of the subsystems. This responsibility includes design, construction, testing, commissioning, participation in operations, planning and execution of machine development, maintenance, diagnosis and repair of faulty components. In general these machine sections will be large contiguous sections of the machine. Typical divisions in the case of a linear $e^+e^-$-collider might be:
- injectors (except damping rings)
- damping rings
- main linacs
- beam delivery/final focus
- (for a 2 beam accelerator) drive beam production system

Similar subsystems can be defined for a $\mu\mu$-collider or a very large hadron collider.

It will probably ease the design, coordination, construction, and operation of the accelerator if the responsibility for large systems is assumed by a group of institutions from one region while smaller systems, such as the positron source or collimation, might be handled by a collaborator with special expertise. To minimize the variety of hardware types to be operated and maintained, collaborators would also be responsible for a particular category of hardware spanning several geographic regions. Responsibility for a hardware subsystem includes all aspects listed above. Examples of such responsibilities could include: (list is not exhaustive)
- control system infrastructure (networks and processors)
- all BPM electronics (or all of a particular type of BPM)
- all power supplies of a particular type
- all RF controls (or all of a particular frequency)
- all vacuum pumps and gauges

In our model we assume that there is a central management which coordinates the design and construction of the machine initially and which later supervises operation and maintenance. The details of such a project organization are outside the scope of this study and remain to be worked out. However, an important input to discuss the technical implications of remote operating is that the following specific tasks are taken care of by a central management. These are:

**Design and standards**
- Definition of overall machine parameters
- Definition of and responsibility for the interfaces between the accelerator sections
- Definition of naming conventions
- Definition of hardware standards and reliability requirements
- Definition and organization of quality control
- Definition of control system standards and interfaces
- Definition of the real-time and off-line database

**Construction**
- Coordination of the construction and installation schedule
- Coordination of the common infrastructure, including
  roads, buildings and tunnels
  power and water distribution
  heating and air conditioning
  cryogenics, pressured air, special services
  site wide communications and networks
  radiation safety systems like shielding, access systems
  fire systems and general safety

**Operation**
- Centrally organized operations
- Planning and coordination of commissioning
- Responsibility for radiation and general safety issues
- Supervision of local maintenance crews, including specialists
- Training of operations and maintenance crews

In order to fulfill these responsibilities, the central management will need technical support and administrative staff.

One particular task of the central management is to produce an overall layout of the accelerator with a consistent set of parameters and performance goals, to make sure that all components of the accelerator will fit together and comply with the requirements for high performance and efficient operation.

We assume that the accelerator is operated under the responsibility of a central operation board, which reports to the central management. This board in coordination with the experiments, is responsible for the mode of operation, the operational parameters, the machine study periods, the interventions, the planning of maintenance periods, the organization of machine operation, and the training of the operations and maintenance crew. This board is therefore the body which oversees all aspects of operations. The local operation crews report to the central board. However, effective safe and high performance operations will depend on information and input from the collaborators who maintain ownership of the accelerator components not only during the design and construction phase but also during commissioning, operation and development periods.

**3.2 Machine Operation**

In the multi-laboratory model, there will be several fully functioning **Control Centers** capable of operating the entire accelerator complex, one at the accelerator site and others at the major collaborating institutions. The operations crew is decentralized and can operate the accelerator from different control rooms around the world. At any given time, the accelerator will be operated from only one of these control centers which has responsibility for all aspects of machine operation. Supporting activities may take place at the other control centers only when specifically authorized by the control center in charge. The current control center handles all accelerator operation including commissioning (to the extent possible), routine operation for physics, machine development studies, ongoing diagnosis, and coordination of maintenance, repairs and interventions. There will also be a control room at the accelerator site which will be needed for some of the initial commissioning and later for complex troubleshooting and specialized machine experiments. Other control rooms may be used for remote diagnosis or machine development experiments in coordination with and under supervision of the active control center. The possibility of control from multiple locations requires a comprehensive system of **tokens** or **permissions**, which are granted by the operators in charge in the current control center.

Control will be handed off between control rooms at whatever intervals are found to be operationally effective. This could be every shift, but might well be in blocks of weeks or months. Operation with multiple control rooms requires very **good documentation** and a mechanism that assures continuity if the operations are handed over to another laboratory. **Electronic logbooks** will be necessary, including a comprehensive log of all commands and events with time stamps, information on the originator, and comments, with a powerful intelligent browser to make the logged information useable for analysis. **Videoconference tools** will be needed for meetings such as shift change, machine status discussions, planning and reporting of machine development. Since local teams in each laboratory perform the machine operations, **standardized training** and operating tools are required. This also means that all control rooms should be identical with a minimum of local dialect and specialization.

In order to help to keep all collaborators around the world well informed and involved, a special effort should be made to make the control center activities visible and transparent when observed from any other laboratory. The machine data will of course be available remotely at any time, but it should also be possible to join in actual control room discussions (as long as efficient and safe operation is assured). Such a '**permanent video conference**' with an open number of participants is not technically possible at this time, but such technology will most likely be developed for other applications and the accelerator community can profit from these developments. Some organizational measures will also be needed for operations meetings to avoid misunderstanding and loss of operation time. For example, a more **formal use of language** with strictly and unambiguously defined elements will be required.

### 3.3 Maintenance

Maintenance is performed by a local crew which, in general, is only responsible for exchanging and handling failed components. The collaborators remain responsible for the components they have built and, in general, these components would be shipped back to them for repair. They must also provide an on-call service for remote troubleshooting. The operations crew currently running the accelerator works with the appropriate experts at their home institutions to diagnose problems. This places stringent constraints on the hardware and its controls interface so that it can be effectively diagnosed from a distance. Local maintenance might be contracted from private industry. Its responsibilities include:
- small repairs
- exchange of faulty components
- assistance to the remote engineer with diagnosis
- shipment of failed components to the responsible institution for repair
- maintenance of a spares inventory

Some tasks such as vacuum system interventions or klystron replacement will require specialized maintenance staff. These highly trained personnel must be available locally to provide a rapid response time. The current operations team has the authority to respond to failures requiring immediate attention. Decisions about planned interventions must be made by the operations board in close collaboration with the laboratory responsible for the particular part of the machine. Interventions are by definition non-routine operations where intense communications are particularly important to avoid misunderstanding, loss of operation time, damage to components, or unnecessary interruption of machine operation. The system of tokens or permissions used to coordinate between the different control centers must have sufficient granularity to allow active remote access to a specific component and to regulate the level of intervention.

### 3.4 Radiation and other Safety Issues

An unresolved question with the proposed model of remote operation is the handling of radiation safety and personnel protection issues. Because of the fact that any accelerator is capable of producing ionizing radiation, its operation must be under strict control in accordance with numerous regulations. In addition to the laws and requirements of the host country and its overseeing government agencies, there are also the internal rules of the partner laboratories which may be even more strict. Current regulations require that there be personnel on site supervising beam operation to guarantee responsibility and accountability. Access to the accelerator housing is another potential problem. While personnel protection systems may be controlled remotely by the operations staff, a hard-wired interlock is usually required to prevent unauthorized access. There are also concerns about the activation of high-power electrical devices and other potentially hazardous systems requiring interlocks and tight control. The task force believes that there exist straightforward technical solutions to ensure the safety and security of these systems. The legal and regulatory issues are a much more difficult challenge which will need careful investigation but they are beyond the scope of the charge to this task force. It is likely that a host country laboratory close to the accelerator site will have to assume responsibility for radiation and other safety issues.

Similarly, unusual events like fires, floods, major accidents, breakdown of vital supplies or general catastrophes will require a **local crisis management** team available on call to provide an effective on-site response. There must be a **formal procedure to transfer responsibilities** to the local crisis management in such instances. This function would also most naturally be provided by the nearby collaborating institutions.

## 4 Evaluation of the Remote Operation Model

In considering the model described in the previous section, the task force tried to evaluate the range of activities compatible with remote operation, the on-site staff required, and some of the sociological issues. The task force concluded that most aspects of machine operation could be performed remotely with only a modest on-site staff for hands-on maintenance and repair. The task force also identified several non-standard approaches which would be advisable to mitigate communications problems in a geographically distributed, international collaboration formed to build and operate an accelerator facility.

### 4.1 Operational Activities

Construction and operation of an accelerator complex requires a diverse range of activities, including commissioning, tune-up, conditioning, normal operation, machine development, maintenance, troubleshooting and repair. The task force did not see any reason why all of these actions could not be performed remotely, at least in principle, with the exception of local maintenance and repair. All modern accelerators have remote controls and do not require access to the component for routine operation. There is no fundamental technical reason why any existing modern accelerator could not be operated remotely. In practice, there are numerous exceptions due to historical reasons, personal preferences and special design choices, but these could easily be eliminated if compatibility with remote operation were a requirement. Older accelerators often relied on analog signals and feedback loops which involved the operator, but modern facilities have already replaced such controls with closed loop digital systems in the interest of efficiency and economy.

Commissioning and tune-up require special modes of operation for the machine components and in some cases, additional remote control signals have to be made available which might not have been required in older installations where local wiring was a convenient option. The experience of any large accelerator complex, however, is that extra effort spent in providing diagnostic information to the control room has proven to be invaluable. The additional effort to build hardware components with remote diagnostics is negligible compared to their total cost. Much of the initial difficulties in accelerator commissioning have typically been caused by insufficient diagnostics. Whenever comprehensive controls and diagnostics have been available in the control room at an early stage of accelerator commissioning, they have facilitated a rather smooth and quick turn-on, as seen at ESRF, PEPII or KEKB. There are many more examples of facilities with insufficient initial diagnostics where progress was unsatisfactory. The conclusion is that any facility with adequate diagnostics and controls for efficient operation could easily be commissioned remotely.

Any large accelerator must also have remote troubleshooting capability, whether the hardware is 30 km or 3000 km from the control room. Many hardware suppliers already offer maintenance packages which allow remote access to the component. Providing this capability in a comprehensive manner should not require any substantial additional cost. The repair of components is usually performed in a location away from the accelerator itself. In a remotely operated facility, this would be done off-site wherever optimum expertise and equipment is available. On the accelerator site, the majority of repairs would involve the exchange of modules. This requires that all components be composed of modules of a reasonable, transportable size which have relatively easy to restore interfaces to the other constituents of the component.

## 4.2 On-site Staff Requirements

From the previous discussion and from recent experience at CERN as DESY as reported in the appendices, only a relatively modest staff appears to be required on site for operations, maintenance or repair. Most of the activities of operation, troubleshooting and failure diagnosis can be performed remotely by off-site personnel, provided sufficient care has been taken to provide remote access. The more complete the diagnostics, the fewer on-site staff will be required. However, even comprehensive remotely accessible diagnostics and troubleshooting systems cannot provide complete coverage of information on everything which can go wrong. Extrapolating from the experience of existing facilities of comparable size, expert intervention on site is required in only about a percent of the failures. If one assumes a rate of 2000 incidents per year (5 per day), there should be not more than 20 occasions where expert help has to be available on site even without relying on further improvements in remote diagnostics and increased modularity and maintenance friendliness of future component design. Distributed across several institutions and laboratories, the additional travel expenses will be negligible. The loss of time may be more severe, and on average, an additional delay of up to 24 hours may be unavoidable a few times per year. Some provision must be made for remote travel on short notice to limit such cases to a minimum.

There must be an on-site maintenance crew which is trained
- to put components out of operation following all required safety rules
- to disassemble a faulty component-module and to replace it by a spare
- to perform simple repairs
- to put the component back into operation
- to release the component for remotely controlled turn-on and setup procedures.

If one were to scale from the HERA experience, one would estimate that the local staff required could be as few as 75 persons. Experience at other laboratories indicates that this number could be higher depending on the details of the hardware.

For efficient operation of the accelerator, regular maintenance is required in addition to the repair of failed components. This work must also be performed by local staff which could be provided by a nearby laboratory or by industrial contractors. The collaborator responsible for the components would plan and manage these efforts under the coordination of the operation board. A small local coordination team of about ten people would be needed to provide the necessary micro-management.

In addition, there must be staff on site for site security, for radiation safety, and for maintenance of infrastructure, buildings and roads. The number of persons needed depends very much on the specific circumstances of the site and the type of accelerator and it is hard to predict a reliable number. In a large laboratory, the staff for these tasks is typically 50-100 people as a rough estimate. In many existing laboratories, there has been a recent tendency to use external contractors for many of these tasks.

In conclusion, the task force estimates that a local staff of about 200 would be required to maintain accelerator and facility operations at a remote site. A part of this staff could potentially come from industrial contractors.

### 4.3 Communications and Sociology

In order to maintain active participation by distant institutions, it is important to keep the distributed collaborators informed about the current status of machine operations. Operation summaries which are continuously updated are currently available for many accelerators. This is a good starting point, but a multi-laboratory facility needs more. Monitors should be available to follow the operations progress and discussions in the active control center. They should easily allow regular 'visits' to the control room as is already good practice in many accelerator laboratories. All operations meetings (shift change, ad hoc meetings for troubleshooting, operation summaries, coordination with experiments, etc.) should be open to collaborators at other institutions. A video conference type of meeting may serve as a model, but present technology has undesirable limitations. The task force expects that growing commercial interest in this sector will promote the needed development.

Spontaneous and informal communications between small groups of people are crucial for distributed laboratories, not only to accomplish work, but also to transmit organizational culture and knowledge, and to maintain the loyalty and good will of the technical staff. Face-to-face communications, whether real or virtual, are the most efficient means of providing such informal and spontaneous connections. Electronic mail and telephone messages also have a role to play, but mostly when only two people are involved. Virtual face-to-face communications can support multi-party conversations, including shared 'blackboards' and computer windows, perhaps using virtual 'rooms' to accommodate specialists with common interests.

In addition to the technical means of communication needed, some organizational measures will have to be taken to avoid misunderstanding and loss of operation time. This would include a **more formal use of language** with strictly and unambiguously defined elements. The components of this formalized language should be accelerator physics terms like *tune*, *damping*, *chromaticity, dispersion, etc.* and operational procedures like *filling*, *correcting*, *tuning*, *centering*, *injecting*, *accumulating*, etc. Formal names are required for accelerator sections, lattice elements, technical components and even buildings. Each of these elements must be defined by a comprehensive description and all communications involving operations must use these *official* terms. This means that a **comprehensive dictionary** has to be written and maintained and the natural reluctance to use it must be overcome.

One concern is that the accelerator teams may cluster in local groups with limited exchange of information. This can be avoided or alleviated by:
- Regular exchange of personnel. This is already good practice in the accelerator community. Mutual visits must be strongly encouraged.
- One or two collaboration meetings per year where the members of the local teams have a chance to meet and discuss with their partners from other teams.
- Common operator training across the collaboration

- A virtual control room which is open to observers from any partner institution at any time. Machine experiments which involve several local teams together should be encouraged.
- Regular exchange and rotation of leadership roles in the project

## 5 Technical Aspects of Remote Operation

### 5.1 Technical Management Aspects

The management structure proposed in the multi-laboratory model was based on the experience from recent large international collaborations for high energy physics and space science experiments. The organizational structure of these collaborations typically includes a mechanism such as a design board for ensuring the coherence of the project. If all of the separate subsystems of the project are not tightly coordinated from the beginning, errors due to miscommunication are inevitable and can be costly to correct later. For an accelerator project as compared with a large detector, the overall performance of the facility depends more strongly on each of the individual subsystems and even tighter control of all aspects of the collaboration effort is required.

Space science projects set a good model of cooperation of several institutions from different countries and industry. They have developed comprehensive standards and procedures on management, design and quality assurance criteria. All of these are well documented in detailed manuals. Such documentation may be a good starting point for discussions on the organization of a large international accelerator project.

The international collaboration would have to develop general management rules and procedures concerning all aspects of the project including costing, bookkeeping and rules for tendering bids in order to have a basis of comparison for the total project effort. There should also be a standard way of reviewing the progress of various subprojects on both technical and management issues. It will be mandatory that these rules be accepted by all collaborators. The task force recognizes that this might be in conflict with the local management culture at the collaborating laboratories, so it appears unavoidable that some change will be required to comply with these demands.

There is also a special problem which deserves consideration but will only be mentioned here: Collaborating laboratories are not always 'independent' organizations but are often part of a larger, usually national organization such as INFN in Italy, Helmholtz Gesellschaft in Germany, URA in the US. There may be no provision for multi-region international collaboration in the existing regulations and procedures of such national science organizations, and a basis for participation in this new type of collaboration would have to be established.

### 5.2 Control System

Modern control systems use a layered approach to communications which should comfortably support remote operation of an accelerator. The entire control system would be implemented locally on site and only the highest level console applications would be accessible from the control centers. High performance facilities rely extensively on automated procedures and closed loop control. These functions often require high speed or high bandwidth and therefore would all be implemented in the local layers of the control system, as would time-critical machine protection algorithms, extensive data logging and execution of routine procedures. The console applications at the control centers would set operating parameters, initiate procedures, monitor performance, diagnose problems and access stored data. These activities require a slower data rate commensurate with human response times, which should not be a problem over any distance on earth. At many sites, the consoles are already 'remote' from the actual control computers and at SLAC, for example, consoles may be run from office or home. The Sloan Digital Sky Survey telescope has a similar controls architecture where the control algorithms are implemented by local processors based on commands which may come from a console located at Fermilab 1200 miles away.

There are different approaches possible for the implementation of the local control and console layers of the control system depending on the requirements of the particular facility. These and other architectural considerations are discussed in more detail below. The networking and bandwidth requirements at the lowest level are independent of whether the control center is local or remote. The requirements for console support are well within the reach of existing technology. The most significant bandwidth demand in a modern control room is for real-time signals which are used for continuous monitoring by the operations staff. In older installations, these are frequently analog signals but with technological advances, essentially all diagnostic devices used to generate such real-time signals now have digital output and network access. Nonetheless, a dedicated high-speed network connection to a remote control room would probably be required to supply sufficient guaranteed bandwidth for this real-time data. It is possible that the new 'quality of service' protocol could eventually provide this functionality over a standard internet connection. Given the rapid rate of development of communications technology, there should be no difficulty in supporting the demands of accelerator operation in 5-10 years.

Most of the existing accelerator control systems use Ethernet LAN technology for data communications. In present facilities, 10Mbit/sec Ethernet technology is sufficient to accommodate the required data rate with an overhead of a factor of ten. The technology for ten times this bandwidth is already available and further development can be anticipated. This should be more than adequate for any future console communication requirements. At the lowest level of closed loop control and machine protection, the need for reliable time-critical data may demand an alternate technology. There is active commercial development ongoing in this area and the solution will depend on the exact requirements of the facility and the available technology when it is constructed. Since these communications are local to the site, the implementation does not impact the feasibility of remote operations.

The evolution of computer hardware and networks has allowed a migration of computing power from large centralized systems to highly distributed systems. This evolution has well matched the march toward larger accelerator complexes. Networks with Gigabit speeds and processors with clock speeds approaching one GHz have pushed far greater control autonomy to lower levels in the controls architecture. These developments favor a 'flat' (non-hierarchical) network structure with intelligent devices which would be directly accessible over the (presumably Ether-) network. Such devices essentially coincide with the current catchword 'Network Appliance', and there will be an immense amount of commercial activity in this direction which will be useful for future projects (for example, power supply manufacturers are already starting to install smart controllers with browser interfaces).

An important consideration in designing a new system is to define the control system to device interface. The interface specification should define the control parameters and diagnostic functions required for each device, as well as the control states used for communication. It is obviously desirable to deal with specific devices in as abstract a fashion as possible. To maintain a clear separation between the local control layer and the remote consoles, one would like to control a power supply in terms of desired current level, duration, and times rather than a continuous string of command parameters. The device should also be able to return diagnostic information needed by the engineers and to monitor internal parameters that define proper operation. Internal device diagnostics should provide early warning to a higher level system when current operation is deviating from previous norms. This information could then be used by the machine beam inhibit system to prevent potentially damaging operation. Smart filtering of alarms is crucial in order to avoid an excess of messages which may easily saturate the system and make it difficult to identify the important information. These are some of the considerations for a system with intelligent devices which minimize the traffic to the remote consoles.

The intelligent device model also implies that the devices be directly on the network rather than hanging on a field-bus below some other device. Traffic can be localized in this structure using 'switches' which forward packets only to the port on which the destination device hangs and whose 'store and forward' capability essentially eliminates Ethernet collisions. This model is now being implemented industrially for factory automation, where one of the strongest motivations for adapting the new technology is the possibility of using standard internet tools for remote diagnostic work.

Important considerations for a large, distributed system are the crucial issues of 'redundancy' and 're-configurability'. The 'Network Appliance' model implies, for a large accelerator, networks with thousands of intelligent local controllers and hundreds of switch/repeater devices. With such numbers failures are certain to occur, and one must consider the tradeoff between the additional cost and complexity for system redundancy, and the downtime required for maintenance. At the network level it will probably be very cost effective to include enough redundancy to permit re-routing if any particular network component fails. At the level of the intelligent controllers the question of redundancy at least deserves serious thought, especially since the price of Ethernet connections should continue to fall.

Where high bandwidth is required for real-time signals, data reduction techniques could also be applied to minimize the amount of data flowing over the global network. Data transmission methods for Digital Television have been developed to minimize the bandwidth of the signal. This is done by periodically sending the basic scene information and then transmitting only the changes to the scene. Similar methods could be used to send reduced byte count messages to the high-level control machines for analysis, data logging and display.

**5.3 Video Communications**

For more than a decade, the technical and sociological issues associated with effective person to person interactions via video telephony have been the subject of academic study. Computer scientists and sociologists have jointly tried several experimental implementations, in trying to draw general conclusions. These studies implicitly assume that spontaneous, informal communication is crucial for distributed organizations "to accomplish work, transmit organizational culture and knowledge, and maintain the loyalty and good will of their members" [2].

In the last decade, video telephony has developed into somewhat more practical video conferencing systems. Nonetheless, it still remains true that "a video link can be barely adequate to promote a shared context and culture to support joint work across two R&D locations, but ... audio and video alone will be insufficient for accomplishing tasks." [3]. Face-to-face communication has traditionally been the primary mechanism through which organizations conduct informal communication (although e-mail now also offers another relatively informal channel). Many of the computer science and sociological evaluations explicitly compare video communications with face-to-face communication. Often it is found that a generic solution is more similar to intentional phone calls than to spontaneous and informal communication. For example, "interviews suggest that the Cruiser system was inadequate because users could not have multi-party conversations and could not share data and other artifacts (e.g., shared blackboards and editors)" [2].

Solutions specific to the needs of astronomers appear to be the most advanced, and to have the most direct relevance to the Global Accelerator Network. In 1992, the US National Science Foundation funded a group of space scientists, computer scientists, and behavioral scientists to launch the Upper Atmospheric Research Collaboratory (UARC). After 6 years, in 1998, the project reported a high level of success in its goal "to provide a distributed community ... with real-time access to remote instrumentation and to provide collaborative tools that would allow them to interact with each other over real-time data (links)" [4].

The UARC collaboration deals not only with multiple sets of remote instrumentation (such as radar telescopes above the arctic circle and satellites), but also with multiple simultaneous "control rooms" (at the University of Michigan, in Norway, in Alaska, and in Russia). "Each user is free to configure their screen as they wish, though there are also tools ... that allow users to share exact copies of windows with their colleagues if they need to coordinate their displays for communications purposes" [5]. The UARC implementation rests heavily on the use of World Wide Web technology and the use of Java applets to ensure interoperability across all platforms. Individuals naturally group themselves into virtual "rooms", which represent functional clusters associated by scientific purpose, as well as places where developers interact. Most of the communication in any particular room is through a multi-party chat facility.

### 5.4 Hardware Requirements

The requirements for the hardware components of a remotely operated accelerator are essentially identical to the usual requirements for a large complex technical facility. The general design criteria are:
- Redundancy of critical parts, if cost considerations allow it
- Avoidance of single point failures and comprehensive failure analysis

- Over-engineering of critical components to increase time between failures
- Standardization of design procedures, quality assurance testing, documentation
- Standardization of components, parts and material wherever technically reasonable
- Avoidance of large temperature gradients and thermal stress, control of humidity and environmental temperature extremes

Specific features connected to remote operation are to foresee:
- High modularity of the components to ease troubleshooting and minimize repair time
- More complete remote diagnostics with access to all critical test and measurement points necessary to reliably diagnose any failure
- Provision for simultaneous operation and observation

A survey of engineers and designers in the major accelerator laboratories indicates that all of these design goals are already incorporated in planning for future accelerators. Due to the large number of components, even with an extremely high mean time between failure, one must expect several breakdown events per day. Even for an accelerator which is integrated into an existing laboratory, comprehensive remote diagnostics are obviously necessary to minimize downtime. This will be one of the crucial technical issues for a large new facility. The mean time between failures has to improve by a factor of 5-10 compared to existing facilities like HERA. This is the real challenge and any additional requirements for remote operation are minor by comparison.

If a device is to be fully diagnosable remotely, it is important that a detailed analysis of the requirements be an integral part of the conceptual design of the component. A quick survey of available industrial products showed that there are already hardware components which meet the requirements. There are several power supply companies which offer remote diagnostics in parallel with operation as an integrated package. There are also many examples of hardware designed and built by the laboratories with remote diagnostics. The modulator for the TESLA/TTF injector linac was developed and built by FERMILAB and is operated at TTF at DESY. The solid state modulator for the NLC main linacs is being designed for full remote diagnosis with no control or monitoring points which require local access.

The conclusion is that the major technical challenges for the hardware of a future accelerator are due to the large number of components and the required reliability and not to the possibility of remote operation and diagnostics. The additional costs for compatibility with remote operation appear negligible.

## Acknowledgements

The authors acknowledge the help and support they received from various individuals inside and outside the accelerator community. We are grateful to the European Space Agency, in particular to Prof. M. Huber who spent significant time discussing the way ESA conducts its projects. Likewise we would like to acknowledge the support of the European Southern Observatory, in particular Dr. M. Ziebell, who explained their experience with remotely operating telescopes. M. Jablonka and B. Aune from CEA Saclay made the experience made in operating the TTF injector from Scaclay available. S. Herb and M. Clausen from DESY and G. Mazzitelli from INFN, Frascati have contributed to the control system sections. C. Pirotte of CERN provided the information on LEP and SPS experience with maintenance and troubleshooting. M. Bieler, J.P. Jensen, M. Staak, F. Mittag, W. Habbe, M. Nagl, and J. Eckolt of DESY contributed information on the technical experience with the HERA complex.

# Appendices

## Appendix A: Experience from SPS and LEP Operations

### A.1 Introduction

The annual exploitation of SPS and LEP can be divided into three phases:
Phase 1: Shutdown for installation, removal and maintenance of equipment (4months)
Phase 2: Startup without and with beam (1month)
Phase 3: Operation (7months)

Clearly for phase 1 a huge on-site presence is needed, needing both specialist technical expertise and full services backup. For phase 2 many interventions are needed (we typically give access daily during this period), so on-site presence is high too. Given the structure of LEP/SPS operation and support, one has to conclude that for both these phases, a large technical infrastructure in needed to support the activities. For phase 3, the necessary on-site support is much less. The evaluation of the necessary minimum support needs to be evaluated. It is the present understanding of the LEP/SPS operation management, that the needed support for phases 1 and 2 lead to the request that a laboratory-like structure must be provided on site.

### A.2 Operation

During normal operation, which is expected to be the largest time slice in any year, things are much more stable. The operators rarely leave the control room, and SPS and LEP are operated under full remote control *when all equipment works*. All real-time aspects, such as synchronization at the ms level, are handled by a stand-alone timing system distributing events around the machines.

So in principle, machine operation could be done from anywhere, with some reservations:
- The LEP operations crew makes use of many analogue signals displayed in the control room, many of which arrive over local links. This could be remedied with modern communications technology.
- Sometimes time-critical actions are necessary, particularly on LEP, such as a frequency shift to save the beams when a RF unit trips. If this has to go through a satellite system, it may not work. However, the main reason for such actions is that the machine is pushed to its limits. Considering SPS operations, which we presently run well within the hardware limitations, there aren't any actions that have to go to the machine at the sub-second level.

However, there are other aspects to machine operation. The PCR is a communication center, where a lot of information is disseminated. It is also a coordination center, where decisions get taken. There is a lot of contact between operations and equipment specialists, between operations and the users, the physicists, between operations and the management, the laboratory directorate. If the control room were situated in one of the major labs, at least part of this communication and coordination could be maintained.

**A.3 Interventions**

High technology equipment breaks down from time to time. *SPS and LEP down time due to equipment faults is typically 10%, so a total of 20days.* Interventions are needed; some of, which can be made remotely, some of which, have to be made on site. Furthermore, some interventions can be made by trained technicians, others need an expert. An analysis of the nature of interventions needed depends on the equipment in question. At SPS/LEP, we have an on-site team who provides first-line intervention for the thousands of power converters used to drive most SPS and LEP equipment. On LEP, where the technology is rather uniform, a large fraction of the interventions, of which there are 1-2 per day, are solved either remotely or by the first-line crew.

At the other end of the spectrum is the LEP RF system and associated cryogenics. I can't see how this system would work without a significant on-site presence of highly trained specialists. Whether or not this has to include the experts who designed and built the equipment is not clear. As mentioned above, there is a high level of communication between operations and the equipment specialists, in this case the RF experts. If the experts are based at the control center, this communication is OK. However, the communication between expert and on-site personnel is severely compromised. This could pose a big problem, since many of the problems encountered (in this system) are not obvious to solve, and need discussion between all parties involved. We have difficulties with this even when all based at the same laboratory!

It is interesting to remember the recent incident at CERN where a transformer caught fire. Once the situation had been brought under control by the fire services, various people were informed and about 2 hours later there were some 50 people at the lab (on a fine Sunday evening). Agreed, not all these people did something useful, but a lot of them were needed:
- Several electrical power specialists, to assess the situation, make safe the equipment.
- Several power supply specialists, one of whom had a good idea enabling SPS to restart quickly, although not quite in the same conditions as before.
- Several safety personnel, to give the go-ahead to restart the machine and to start investigations into any side effects of the fire (such as environmental hazards).
- Several experienced operations personnel to modify machine settings, in collaboration with the power supply specialists, to adapt to the new configuration.

All this was done in a matter of hours. Dealing with this kind of situation at a remote facility would have taken days if not weeks. On the other hand it doesn't happen often.

**A.4 Conclusions**

A significant infra structure will be needed close to the facility, fully staffed for many months per year, for installation/removal/maintenance of equipment.

An on-site presence will be needed during operation, for intervention on faulty equipment that cannot be fixed remotely. This does not strike me as a very enjoyable job, stuck out in the middle of nowhere, rather like working on an oilrig.

To have any hope of running a major facility remotely, it will have to be operated well within its design limitations. The power converter operation is a good example of what can be done. If we push things like we do now on LEP RF, it will not work.

If it is possible at all, we will have to expect a drop in productivity due to the generally lower level of efficiency of interventions.

We can try to learn from similar experiences in related and other fields.

## Appendix B: First Line intervention on the power converters for LEP, SPS and the EA

**B.1 Summary**

Looking just at LEP, which is most representative of a future machine, the statistics are impressive. In a year they make over a hundred called-for interventions, solve 85% of them directly, a further 10% after telephone conversation with a specialist. Of the remaining 11, 8 are solved by the contractor's specialist, and only 3 needed CER specialist intervention. The same team also undertakes repairs of exchanged units and does preventive maintenance during the shutdown. So a system like this can be run and maintained by a small on-site team, with very little specialist intervention. This does not come for free. The system has to be designed with this in mind (highly modular, standard equipment) and you need fully trained people to run it.

There is also the psychological side of having a remote team doing this kind of work, which could be at a remote site. Persons in charge of maintenance and operating felt that the turnover would be very high! This brought up the idea that it may be much better to have the operations team on site. So then we are looking at something like 50+ people to run the machine, fix problems, maintain equipment etc. This would provide a more reasonable working environment than just having the equipment fixers on site. However, it weakens the link between operations and the specialists who are based at the labs, and means that turnover would be high in operations too!

**B.2 Accelerator reliability and the work of the maintenance and repair team.**

LEP and SPS maintenance and intervention is performed by a First Line Crew, which has been trained to perform work on almost all hardware systems. Such a crew is possibly needed for a remote facility which makes the LEP/SPS experience very valuable.

**Lep**

LEP Power Supply fault statistics:
Number of power converters: 825
Number of incidents per day: 0.5

About 85% of the problems can be solved by the first line team (FL). These include exchange of whole units and in situ repairs. In 15% percent of the cases, experts need to be contacted. In 10% percent of the cases, the problems could be solved by telephone conversation. In 5% percent of the cases, experts were needed one site. In only 3% of the cases, CERN experts were needed on site to fix a problem. The latter corresponds to 3 times in a year of operation (excluding shut downs and start-up). These numbers do not include interventions and diagnostics made by the experts in charge during normal working hours. These numbers also do not include scheduled maintenance, which is planned and performed by contractors.

It should be noted that future remote diagnostics could be improved considerably by using modern communication techniques. This is expected to reduce the number of necessary interventions by experts. It should be further noted that quite a few incidents are not caused by malfunction of the power supplies but are caused by ground faults, which preferably occur after maintenance work in the accelerator tunnel. Most of the expert interventions were necessary to solve problems with big power supplies (main dipole and quadrupole circuits). A more modular design of such units could further reduce the number of interventions.

**SPS**

SPS Power Supply fault statistics:
Number of power converters: 577
Number of incidents: 0.360 per day

The statistics is basically the same with a slightly higher percentage of expert interventions. These reasons for this are that SPS components are less standardized, documentation is not quite as good, and the technical realization of components is not as friendly for exchange of faulty components. This means that the number of in situ repairs is larger.

**Zones**

Number of power supplies 485
Number of incidents 0.72 per day

The fault rate is much higher for the beam lines due to old equipment, but the number of specialist interventions are small. Most of the problems can be easily identified. Most of the problems are simply contact problems. A failure of a power supply is also less critical as in case of an accelerator so that there is less urgency to solve a problem quickly. The fault statistics is better in the period following a large shutdown where systematic maintenance takes place.

**B.3 Conclusions**

The presence of a crew of technicians on site is necessary. But LEP experience shows, that a well trained crew with a good knowledge of the machine (which requires one year of training) can solve 85% of all the problems without the help of the experts and 95% percent of the cases require no more than aid via telephone. The technicians must have a good overview of all hardware systems and shouldn't be too specialized. Preventive maintenance helps to reduce incidents during operation. Fixing problems is eased considerably by: Spares stored in situ, modular concept of design, standardization of equipment, well established diagnostics procedures, well organized stockholding for spares, repair-friendly design, careful testing of repaired equipment.

It is most essential to provide comprehensive remote diagnostics (this is true not only for the power supply systems). The remote diagnostics must be user friendly to be useful for the first line technicians as well. New communication technology supports remote diagnostics and to certain degree remote interventions (setting of parameters, system configurations etc) It is important to provide clear interfaces and borders between subsystems. There should be regular consultations and exchange of experiences between maintenance and repair crew and experts.

## Appendix C: HERA Experience

**C.1 Power Supply System**

The HERA Power supply system is a vast system with more 1200 Power converters. Most of them are chopper-type supplies (about 800), the rest are thyristor-type supplies. Between January and July of 2000, there were 42 incidents of power supply failure which caused a beam loss and required intervention. This is a very good number which corresponds to 400,000 hours of operation between failures. All these interventions were performed by a crew of technicians from the power supply group. This group provides a 24h 7-day service. The persons in charge are able to resolve the problem without the aid of experts (one or two exceptions in 2000). However, it should be noticed, that the experts themselves are also part of the emergency crew. Without these experts (20% of the crew) on duty, the number of problems which cannot be solved without external expert help would go up slightly to at most 4 incidents per year.

Most of the problems in the past had to do with new designs which have been improved in operation. Problematic were electro-mechanic polarity switches used in most corrector supplies, transistor switches in power chopper supplies, non-tight water cooling hoses, loose contacts, bad connectors. The situation improved drastically after the decision was made to exchange faulty components on the first occurrence of a fault. A more difficult problem to handle is ground faults. To find ground faults quickly in a vast system often require expert knowledge on the magnet system. On the other hand, a ground fault is very often due to an unexpected usually "trivial" reason. Preventive maintenance during shutdowns is another important factor in reducing the number of incidents.

Most of the power supply failures in the HERA complex can be sufficiently well diagnosed remotely. Nevertheless, there is a policy that certain classes of errors must be reset at the power supply after a visual inspection. This could be avoided by doubling or tripling the information remotely available and using modern communication technology. Some of the modern large power supplies (for example the proton main circuit 8kA, 500V) have a fully computerized interface which allows complete diagnostics. Most of the vendors of large power supplies today offer a remote diagnostics and troubleshooting service.

Taking these options into account one may conclude that the HERA power supply system could with some moderate effort be operated fully remotely. Interventions, exchange and in situ repairs could be carried out by a non-expert crew with a general technical training. The modest effort required would include a more complete survey of power converter by plc technology, somewhat more modular design of the components, better documentation using internet tools for example, portable maintenance and troubleshooting computers which can be connected to a database containing all necessary information which is carefully organized according to the need of a non-expert crew.

**C.2 RF Systems**

The rf system of HERA consists of 8 stations, each driven by 1.5MW 500MW-douple klystron. There are a total of 85 room temperature and 16 super-conducting multi-cell cavities. The system runs very reliably at this time. Only 52 incidents occurred in the year 2000 in this big system. The system has a power overhead of about 40% which turns out to be an important factor in reliable and smooth operating.

Preventive maintenance is an important ingredient in reliable RF operation. All the technical interlock functions are checked regularly by artificially initiating a fault situation. Parts like filters, fans, seals, sensors are exchanged on a regular basis. High Voltage carrying elements of the modulators which have large surfaces exposed to air need regular cleaning and polishing. Electronics after an initial phase of troubleshooting (which can be several years in some cases) turns out to be robust. This preventive maintenance could be performed without problems by a non-expert maintenance crew. To allow complete remote checkout of all the interlock systems would be possible but would be quite expensive (add in heaters and switches and valves to remotely simulate fault-situations). However, it should be easy to monitor the test remotely and do the evaluation remotely.

As far as diagnostics and troubleshooting during operation are concerned, the experience with the HERA RF seem very remote-compatible. Most of the incidents are due to hidden faults, or due to a non-faulty situation which nevertheless triggers an interlock condition. Usually these faults occur under high power running conditions and are often not reproducible. Therefore, the only way to reduce these faults is to have a complete recording of all the relevant parameters during a fault (transient recorders, circular buffers) and comprehensive data logging. Thus the conclusion is that in order to be able to find the relevant problems at all, a big effort in diagnostics is necessary. The remote diagnostics capability thus does not cost any extra effort.

Repair of faulty components can be performed in most cases by a non-expert crew given improved documentation and good training. RF related problems do usually require expert intervention. Not all of them can be carried out remotely, but these cases are quite rare (about 1-2 incidents per year)

### C.3 RF High Voltage Supply System

Remote maintenance is no problem for all electrical features of the system. Visual inspections are necessary to check contacts and mechanical stability of the system. High voltage cables, in particular the end cap are a neuralgic point in this system. These tasks do not need to be performed by the laboratory expert but could be performed by a non-expert crew. Regular checks are necessary and require experience to catch a problem before it causes loss of operating time. Repairs on high voltage systems like transformers, switches, cables are usually performed by external highly specialized companies.

### C.4 Quench Protection System

This system is the heart of the HERA safety systems. It collects information on all the subsystems like superconducting magnets, power supplies, rf systems, beam loss and position monitors. It continuously evaluates the system and in case of an incident makes a decision to dump the beam, to ramp down the main power circuit or to dump the magnetic stored energy quickly in dump resistors. This system needed quite some debugging time. The biggest problem turned out to be a bad contact (due to poor cable manufacture). The system is controlled by PLC. At each incident a complete snapshot of the system is stored which allows troubleshooting of the system almost completely remotely. Initial debugging however quite frequently required expert presence at the location. In 2000, the system had only 6 faults during operation. They were diagnosed remotely and required expert presence on only one occasion.

The experts on this system would like to emphasis the following points:
1) It is important that the design include comprehensive remote diagnostics, redundancy, and modularity.
2) The use of PLC systems has been very successful. They are quite flexible for accommodating new subsystems. They also support the feature of a freeze buffer which provides a complete snapshot of the system in case of an event or error. This allows detection of the sequence of occurrences, in particular to detect the "first error". A digital transient recorder for storing events with timestamps on a circular buffer is very helpful for "post-mortem"-analysis.
3) Remote booting capability for the distributed computer system is very essential.
4) Very important is regular maintenance and tests of components which are not frequently used (for example quench heater systems).

A complete loss of direct contact to the hardware is hard to imagine for most of the experts. To maintain the skill and expertise there must be the possibility for occasional local inspection and tests on site (once per year?).

## C.5 Conclusion

It required a lot of effort to achieve the present level of reliability of the HERA components. At present, there are about 1.5 incidents per day. 50% of the faults can be just reset from the control by the operating crew. About 20 % of the remaining incidents need some closer look remotely by more expert staff. In about 25% of the cases, some action is required in situ, which is taken care of by the operators. In 5% of the cases remote expert help is necessary. In about 1% of the cases, an expert is called in to fix the problem. Many problems however are resolved or worked around only temporarily in this way. Expert actions are required on maintenance days to fix the problems completely.

For remote operation, one can draw the following conclusions: HERA experience suggests that ii would be possible to operate a facility like this completely remotely. A local group of technicians would be necessary however to do simple repairs, to exchange faulty components, to assist in remote diagnostics and to do maintenance. In order to provide efficient operation, a large effort (compared to what has been done for HERA) has to be made to provide adequate information, to improve the remote diagnostic capability, to systematically train the operation and troubleshooting crew. The occasions where this is not enough are expected to rather infrequent (2-3 times per year).

During the initial debugging phase of the facility however, many unforeseen faults and errors occur and design deficiencies are revealed. The experts have a hard time to imagine that this could be resolved completely remotely. This means that during commissioning a longer stay of the experts on site may be unavoidable. There must be an infrastructure on site to accommodate experts during this period.

# Appendix D: Experience at the European Space Agency

**D.1 Introduction**

On November 27 2000, task force members visited the European Space Agency's (ESA) technical centre ESTEC in the Netherlands. ESTEC management and technical staff was very open and helpful to allow a look into ESA/ESTEC management procedures, experience in the design of equipment for remote operation in space, and operational experience. A synopsis of this information is given in the following section. The ESA is the European Space Agency. It performs and supports experiments which are carried out in space. ESA launches satellites which are in part provided by partners from universities and other science institutions. The experimental set-ups are often produced with a strong contribution from industry. Sometimes industry takes the position of the prime contractor which integrates contributions from subcontractors and institutes. The strong interaction between ESA/ESTEC, its scientific partner institutions and industrial partners make ESA a good model for remote operation of a future accelerator facility, in particular with respect to project configuration and project management. The scientific partners of ESA have sometimes the character of users of the ESA facilities which bring their experimental set-ups and equipment in space, which keeps and controls it there, and which provides the links for transmission of experimental data. This has some resemblance to the relationship of an accelerator laboratory and its user community.

**D.2 Management issues**

ESA consists of a number of institutions which are distributed all over Europe. In order to perform its projects efficiently, it has developed strict and formal management rules and procedures. A management manual is available which describes all management processes at any stage of a project in a formalized way, specifying input, output and the procedure. It would be interesting to compare these management rules to the corresponding procedures which are practiced, for example, in DOE laboratories or at CERN. These rules describe and specify:

- System specification requirement, the rules of how to write a product specification
- Product insurance requirement, the rules of how to accompany the production process
- Verification requirements
- Management requirements
- External interface requirements

The management procedures also contain elements like configuration control, change control, non-conformance control. These management rules are also mandatory for the ESA/ESTEC collaborators and contractors to obtain good transparency and tractability of the whole project.

In the discussions with ESA management staff it appeared obvious that a world wide accelerator collaboration has to develop similar types of management rules which are obeyed by all collaborating institutions. This is necessary to keep the schedule, the costs and the quality of the components under control. In particular, the central management is responsible for the "top specification" of the product. As established in ESA projects management instruments like change control are needed to support and maintain this specification process. The central management, assisted by a central steering committee, must be put in the position to execute inter-phase control and surveillance of the production process. This means that the collaborators are not completely free in their management, and technical choices. The central management must have the authority to allocate freely the considerable funds necessary for the task of integrating the components delivered by the contractors and collaborators.

**D.3 Design Process**

The experimental set-ups and the technical support of ESA projects are usually out of reach once in operation. Therefore a very high quality standard is necessary which is one of the important objectives of the design process. One way of achieving high quality of the design are standards. There are standards in production procedures, material choices and standards in training of technical personal.

A large fraction of the management rules and procedures are about specifications. Specifications are written in a formal way. There are many rules such as: "Never use several specifications in one sentence." Quite helpful is a formalized component alert system which is accessible to all the contributors which provides a well-maintained list of non-satisfactory products which must be avoided in the design.

Remote failure analysis is an important design input, which must be taken into account starting from the beginning of the design process. There are no standards on remote diagnostics. Case to case decisions must be made.

It is ESA experience that a formal reliability analysis based on the statistical reliability of single components is not so helpful. A "statistical failure" of equipment is apparently very rare. Most of the errors are produced by hidden design errors or circumstantial elements which have not been taken care of properly in the design. (This experience is similar to what is seen with accelerators.) What has been very important is the impact of failure analysis. It is very important to understand what a failure of a component or a part of the set-up means for the functioning of the rest of the system. This of course has also been experienced with accelerator equipment.

There are other elements in the process to reach a good design, which are less relevant to the accelerator world. This is extensive testing under extreme conditions (mechanical, temperature, acoustical, etc) and redundancy. While full redundancy may not be feasible for accelerator components because of costs when the number of components becomes large, it is an important input that redundant equipment should be physically well separated. This design rule is also relevant for accelerators.

An important design goal is to avoid single point failures. Sometimes, this may be unavoidable. In this cases it is very important however to identify single failure points. ESA doesn't have specific design choices aside from things like equipping connectors with double pins and using only coated conductors. Specific design decisions are produced moreover made case by case.

**D.4 Operational Experience**

The data rate for the transmission of system data and experimental data is limited (100kbit /sec) and in competition. Therefore, a low but steady flow of system data is transferred and analyzed continuously. This data logging is called "house keeping" data. It is used for trend checking to catch a failure at as early a stage as possible. These data are analyzed in the control center. Remote reset and switch over to redundant equipment are possible actions as a consequence of this analysis. Remote interference by experts far from the control center is a standard procedure. ESA staff was quite interested in this context to discuss the communication tools used in the accelerator community.

**D.5 Conclusions**

ESA experience might be useful to study in the process of modeling remote operations. Quite interesting are the management and design procedures in this context.

Report of the ICFA Taskforce on Technical Aspects of a Global Accelerator Network## Appendix E: Remote Operation of Telescopes by the European Southern Observatory ESO

In choosing the location for large telescopes, the needs for optimum geographic conditions for observation cannot be compromised. For this reason, large telescopes are often constructed at inhospitable mountain sites close to the equator, usually far from any urban infrastructure. Quite consequently, astronomers started already in the 1980's to operate telescopes remotely and much experience in remote operations has been gained.

The European Southern Observatory (ESO) has operated for two decades large telescope facilities in the Andes Mountains in Chile.

The ESO is a European institution which builds and operates telescopes and makes them available to users from smaller institutions, typically from University type institutes. In this respect there is some resemblance to a synchrotron light source laboratory with a large user community. Admission and access to the experimental facilities are granted by using similar approval procedures as the ones used in the accelerator community.

The headquarter of ESO is located at Garching near Munich in Germany. Most of the scientific and expert staff are situated there. There are about 300 staff members stationed in Garching. ESO has two main remote sites in Chile. There is first the site located in La Silla at an altitude of in 2400m, where the 3.6m CAT and NTT telescopes are situated. The second site in is Paranal at 2600m altitude which hosts the VLT facility with four large 8m telescopes. The remote site staff is 100 and 100 persons respectively. The staff consists mostly of non-scientist, technician-level employees.

Telescopes are complicated mechanical machines. The large, multi-ton mirror systems are positioned and stabilized with an accuracy of 0.1 arc seconds. These are controlled by typically 50 large electrical motors and 400 small motors. These motors are controlled by sophisticated, complicated, automated, nested control loops. Information is feed in by encoder systems with a 1nm precision. Typical time constants of the regulation loops are 12 ms. In addition there are the various detector and observation systems which also provide feedback for the positioning system. The costs of such a facility not counting building and site infrastructure is about 30Million DM. This complex technical system is thus comparable to a small accelerator.

The availability of a telescope facility is of this kind is about 85% of the usable scheduled operating time.

The control system of these telescopes is three-layered. A local area network connects the intelligent front-end controllers and local intelligent control loops of the positioning systems with a middle-layer workstation that coordinates the various tasks. The third layer is the operating level. It is outside the high bandwidth control loops and accommodates the input and output for the human interface. The communication technique is based on Ethernet. For precise timing, an analog system is used in parallel. At the hardware near level, industrial standards such as VME with a VX-works operating system is widely used. This control architecture resembles very much the ones of modern accelerators.

Due to the remote site, the CAT and NTT telescopes were operated for many years from Garching using remote access to the local computer network. The remote access was limited strictly to the third, operator-accessible level of the control system. The far distance networking was eventually based on a laboratory owned satellite 12-14GHz (C band) connection which offer a data rate of 700kbit s$^{-1}$ which is more than sufficient for operating and acquisition of experimental data. The signal duration from Garching to Chile is 450ms, which is; though somewhat slower than ground cable connections, still sufficiently fast for videoconference transmissions. Unfortunately, for larger distances, the transmission time would be increased considerably due to the then necessary deviation via intermediate ground stations. This approach not only provided a good data rate, it proved to be the most cost effective way of data communication and provided on top the best conditions operational safety and stability as well.

Remote operations from Garching were performed without any technical problems. A large fraction of troubleshooting could also be performed from afar. However, repairs and tuning on the complex mechanical systems of the telescope were usually performed by experts on site. The very delicate dynamics of these systems, with the potential of causing much damage if improperly adjusted, seemed to be prohibitive for completely remote handling or repairs by non-expert staff. For these reasons, the local experts which were initially located at the city of Santiago or Garching have been relocated on site in order to be more effective. The time experts devoted to troubleshooting and repair work dropped from 30% to 5% by this measure which seemed to justify the inconvenience of staying for longer periods at the remote site. With an expert crew on site, remote operations lost its attractivity. While the CAT telescope was during its whole lifecycle, operated from Garching, the NTT telescope was operated locally after its control system was modernized. The site in Garching did not follow the modernization process and the local and remote site became incompatible.

Commissioning of new installation is for the reasons mentioned above also performed by experts on site. ESO technical managers insist that emergency stops and similar safety features must be hardwired and it is not safe enough to transmit them from afar.

A new international project named ALMA is presently under discussion. It is a large international collaboration of the kind envisioned for a future large accelerator project. It consists of a vast system of 64 antennas would be installed at the Chanjandor plateau in Chile at 5200m altitude. It is obvious that it will not be possible to have a large number of staff permanently stationed at such a high altitude. The control center of this facility is therefore planned some 30km away in 2000m altitude. Still present plans call for experts on site for the commissioning of the complex. Nevertheless in view of the difficulties of working in such inhospitable environments, remote operation is considered as a very attractive option. It will be very interesting for the accelerator community to observe the progress of this project which is quite similar to a future accelerator project since it includes international collaboration, remote sites, and large, complex, and challenging technical systems.

ESO has a large experience in videoconferencing. Usually there are three videoconferences per week. ESO is in the process of modernizing its systems. Modern systems with 3 channels a 64kBitsec$^{-1}$ or more constitute a large step forward in bandwidth and quality in sound and picture.

## Appendix F: Remote Operation of the Sloan Digital Sky Survey Telescope

The Sloan Digital Sky Survey (SDSS) is a project substantially funded by the Alfred P. Sloan Foundation. The project will create a comprehensive visible digital photo-metric map of half of the northern sky to very faint magnitudes. This project has constructed a 2.5-meter telescope located at Apache Point Observatory in southern New Mexico. The telescope employs the first imaging camera to use large-area CCD detectors. A unique feature of this telescope is the fact that the telescope is not contained in the more traditional "dome" used by other telescopes. This telescope has a service building, which is rolled away prior to the beginning of telescope operations. The telescope employs a Wind Baffle system, which is independent of the telescope to shield it from wind buffeting. This system is servo driven and is slaved to the telescope motion using two linear voltage differential transformers (LVDT). The telescope itself uses three axis control, altitude, azimuth and rotator, for pointing.

The axis control, developed by Fermi National Accelerator Laboratory, is based on a VME bus system. This system uses two microprocessors and a variety of support interface cards. The Motion Control Processor (MCP) receives position, velocity, and time (pvt) instructions from a host Telescope Control Computer (TCC) located in the control room. This room is located on site, but remote from the telescope location. The pvt command instructs the telescope MCP to have the telescope at a certain position at a certain velocity at a certain time. The MCP maps the pvt information into motion commands for the three controlled axes. These commands are sent to the servo controller card. This card uses a Digital Signal Processor (DSP) to control the servo loops on each axis.

A second processor card, the Telescope Performance Monitor, is used to track and log numerous parameters during telescope operation. Various parts of the servo systems are logged with this data. Servo output voltages, current, command position, actual position, position error, and servo amplifier status are among the information logged at a 20Hz rate. The telescope has an extensive Safety Interlock System used both for personnel safety and equipment safety. Data from this system is fed to the TPM to allow easy diagnosis of interlock conditions from the control room.

If one examines the underlying architecture of the telescope controls, you quickly find that the local control system does not know where the control room is physically located. Data communications between the TCC and MCP/TPM are handled by Ethernet. It makes little difference to the system if the commands come from the control room at Apache Point or from a remote site. In the course of troubleshooting any servo problem, the system engineer at Fermilab, located 1200 miles away, can connect to the MCP and issue motion commands remotely. Equipment parameters are returned via a TPM link or directly from the MCP via built in diagnostic tasks. Equipment and personnel safety are still handled by the Safety System on site. On-site personnel are used to replace components and ship them back to Fermilab for repair if on-site repair cannot be done.

It is important to note that due to various instruments used by the SDSS survey the presence of observers is still required to perform the instrument changes and other maintenance items. The purpose of this paper is to show that remote operation of a complex device is possible and has been accomplished.

## Appendix G: Control of the TTF linac from Saclay

As part of an international collaboration, the CEA laboratory in Saclay, and the LAL laboratory at Orsay in France have provided the injector for the Tesla Test Facility (TTF). The injector consists of a thermionic gun, an RF buncher structure and a superconducting capture cavity. It includes also the connecting beam lines, beam instrumentation, beam analysis channel and the RF power sources. It is quite a complex hardware system and due to the high beam intensity, the machine protection aspects in the operation of this facility are crucial.

The control system of the injector complex is designed for remote access. Any physicist or engineer participating in the TTF collaboration can run all the application programs which are used to control the injector at DESY from their office using a standard X-terminal and internet connection. This allows the machine to be operated remotely with good reliability, and the response appears "instantaneous". The only control on such access is knowledge of the common password. So far, no incident nor unwanted action on the machine has occured.

This feature has been used for several years by Saclay physicists for the control and the maintenance of the Capture Cavity RF system. The capture cavity (CC) is a standard TTF superconducting cavity housed in a separate cryostat and powered by a separate klystron and modulator. Placed right after the RF gun, it determines to a large extent the beam quality, the transverse and longitudinal emittances, and the stability of these parameters both within the macro-pulse and long term pulse to pulse. The RF is pulsed (100 to 800 µs, 1 to 10 Hz) and must be regulated in phase and amplitude within narrow limits (0.2° and $5.10^{-4}$). The cavity frequency shifts during the RF pulse due to Lorenz forces and because of the high Q factor and particular geometry of the cavities. This requires a complex regulation system. Four feedback loops have to be optimised, closed or opened at different adjustable times, and a tuning system has to be operated both coarsely (after a cool down) or very finely before beam operations. The klystron and modulator are more conventional devices but still, as in any accelerator, require experts for their maintenance.

TTF physicists, on the other hand, do not have expertise in all of the many complex devices installed. As a result, it was soon found desirable to have the permanent assistance of Saclay physicists in the management of the CC RF, which was possible through the Internet connection. Since then, it is a frequent occurrence that either an operator in the control room at DESY asks a Saclay colleague to make necessary adjustments or one in Saclay calls the control room for permission to perform corrections or tests. Both operators can also work together with only telephone contact. Signals reconstructed from fast sampling as well as long term recorded data are available to the Saclay operators. Software modifications can be downloaded and tested from Saclay.

In principle, the whole linac can also be controlled remotely, except for the real-time video images, which are too slow for on-line utilisation. There has not yet been an attempt to adjust all the beam transport through the linac or to perform a complete experiment remotley but it could certainly be envisaged. The on-line logbook at DESY has proved very useful for the Saclay physicists in understanding ongoing operations.

Overall, the Internet connection to the TTF control room has now become a permanent, very convenient and satisfactory feature.